\documentclass[11pt]{article}
\usepackage{graphicx}
\usepackage{amsmath,amsfonts,amssymb}
\usepackage{color}
\usepackage{bbold}
\usepackage{braket, simpler-wick}
\allowdisplaybreaks
\usepackage{charter}
\def\sloppy{\tolerance=100000\hfuzz=\maxdimen\vfuzz=\maxdimen}
\allowdisplaybreaks
\def \beq  {\begin{equation}}
\def \eeq  {\end{equation}}
\def \beqar {\begin{eqnarray}}
\def \eeqar {\end{eqnarray}}
\def\sqr#1#2{{\vcenter{\vbox{\hrule height.#2pt
\hbox{\vrule width.#2pt height#1pt \kern#1pt
\vrule width.#2pt}\hrule height.#2pt}}}}

\mathchardef\mhyphen="2D

\def\la {{\langle}}
\def\ra {{\rangle}}
\def\vx {{\vec x}}
\def\vy {{\vec y}}
\def\vk {{\vec k}}

\def\Tr {{\rm Tr}}

\def\vk {\vec{k}}
\def\vp {{\vec p}}

\def\vx {{\vec x}}

\def\vy{\vec{y}}
\def\vv {\vec{v}}

\def\del {\partial}

\def\D {{\cal D}}

\def\E {{\cal E}}

\def\V {{\cal V}}

\def\half{\textstyle{1\over 2}}
\def\quarter {\textstyle{1\over 4}}

\begin{document}
\def \CMP {{Commun. Math. Phys.}}
\def \PRL {{Phys. Rev. Lett.}}
\def \PL {{Phys. Lett.}}
\def \NPBProc {{Nucl. Phys. B (Proc. Suppl.)}}
\def \NP {{Nucl. Phys.}}
\def \RMP {{Rev. Mod. Phys.}}
\def \JGP {{J. Geom. Phys.}}
\def \CQG {{Class. Quant. Grav.}}
\def \MPL {{Mod. Phys. Lett.}}
\def \IJMP {{ Int. J. Mod. Phys.}}
\def \JHEP {{JHEP}}
\def \PR {{Phys. Rev.}}
\def \JMP {{J. Math. Phys.}}
\def \GRG{{Gen. Rel. Grav.}}
\begin{titlepage}
\null\vspace{-62pt} \pagestyle{empty}
\begin{center}
\rightline{CCNY-HEP-18-03}
\rightline{April 2018}
\vspace{0.7truein} {\Large\bfseries
An Action for the Infrared Regime of Gauge Theories}
\vskip .05in
 {\Large\bfseries
and the Problem of Color Transformations}
\vspace{16pt}
{\Large \bfseries  ~}\\
{\Large\bfseries ~}\\
 {\large\sc A.P. Balachandran$^1$} and {\large\sc V.P. Nair$^2$}\\
\vskip .2in
{\itshape $^1$Physics Department, Syracuse University\\
Syracuse, NY 13244-1130}\\
\vskip .1in
{\itshape $^2$Physics Department,
City College of the CUNY\\
New York, NY 10031}\\
\vskip .1in
\begin{tabular}{r l}
E-mail:&\!\!\!{\fontfamily{cmtt}\fontsize{11pt}{15pt}\selectfont balachandran38@gmail.com}\\
&\!\!\!{\fontfamily{cmtt}\fontsize{11pt}{15pt}\selectfont vpnair@ccny.cuny.edu}
\end{tabular}

\vspace{.8in}
\centerline{\large\bf Abstract}
\end{center}
It has been known for a while that there is spontaneous breaking of Lorentz symmetry
in the nonzero charged sectors
of quantum electrodynamics due to the infrared problem of soft photons.
More recently, it has also been suggested that similar results hold for
color transformations in a nonabelian gauge theory. Here we show that an action
where a diffeomorphism  has been carried out for the part describing
hard gauge particles and matter
fields can be used to analyze these issues. In addition to rederiving old results
in this formalism,
we also show that color transformations cannot be unitarily implemented
on perturbative gluon states if gluon fields of arbitrarily low energy are allowed.
Implications for confinement and mass gap are briefly commented upon.

\end{titlepage}

\pagestyle{plain} \setcounter{page}{2}

\section{Introduction}

It has long been recognized that the existence of infrared divergences
in a gauge theory leads to subtleties in the definition of charged states
\cite{{blochN}, {weinberg}}.
Charged particles are effectively accompanied by a cloud
of soft photons.
The analysis of scattering amplitudes shows that the charged states
become non-Fock coherent states which may be defined by the action of
a ``dressing factor" on the Fock vacuum of soft photons
\cite{{blochN}, {chung}, {FadKul}}.
The effect of this dressing on the S-matrix, specifically
how it helps to factor out
the infrared divergences, has
become standard textbook material by now \cite{weinberg}. Nevertheless,
this subject has seen a recent revival of interest
related to 
the role of asymptotic symmetries and their consequences such as soft photon theorems
in electrodynamics \cite{othersQED}.
Another curious feature regarding the non-Fock coherent states
which emerged from 
the extensive analysis carried out over many years
 is the spontaneous breaking of Lorentz symmetry in the nonzero 
charge sectors of the theory \cite{{buchh}, {morchio}}.
The dressing factor which leads to the coherent
states is generated by the asymptotic soft radiation field
associated to charged particles and
is characterized by
a time-like vector $p_\mu$. The overlap of such coherent  states for different choices
of this vector, i.e., for $p_\mu$ and $p'_\mu$ (with $p'_\mu \neq p_\mu$) is zero.
 In other words, if 
we think of $p'_\mu$ as a Lorentz transform of $p_\mu$, the corresponding 
transformation of the states
cannot be unitarily implemented.
While this spontaneous breaking must be taken account of in the theoretical set-up of quantum
electrodynamics, its physical implications are less obvious.
There is no such breaking in the sectors with zero net charge; since this 
sector is what is adequate for almost all practical situations (such as laboratory
experiments on scattering),
there is no easily obtainable observable consequence.

Given this situation, one possibility is to see if operators which are sensitive to this issue
can be incorporated into the theory \cite{bal1}. For example, in electrodynamics, one such 
operator is given by $U (\theta ) = \exp( i Q (\theta ))$ where
\beq
Q(\theta) = \int d^3x~ ( - \nabla\theta \cdot {\vec E} + \theta \, j_0 )
\label{i1}
\eeq
If $\theta$ is a smooth function on the spatial manifold
${\mathbb R}^3$ which vanishes at spatial 
infinity, then $Q(\theta )$ is just the Gauss law and hence it will vanish on
physical states.
For a function $\theta$ which becomes a constant (say $\theta_\infty$) at spatial infinity,
$Q (\theta )$ becomes $\theta_\infty$ times the charge operator,
while
for those functions $\theta$ which
tend to nontrivial functions on the two-sphere at spatial infinity, $Q(\theta)$
furnish a set of operators sensitive to the asymptotic behavior of the fields.
One can attempt modifying the theory in a way which depends on these; for example, a 
modified mass term $m \rightarrow m ( U(\theta ) + U^\dagger (\theta ))$ \cite{{bal1}, {bal2}}.
Interference effects may be another way to detect consequences of the
infrared dressing factor \cite{sem}.

Another set of questions arises from the familiar consequences of spontaneous 
symmetry breaking.
Is there a Goldstone mode one can identify? Also, recall that while the Goldstone theorem
and the Higgs mechanism can be explained in terms of correlators of operators, there is
a simple action-based description which captures the essence of the phenomena.
One can ask if there is a similar description in the present case.
Finally, in a nonabelian gauge theory, where the charge algebra is nonabelian,
the dressing factor is
applied to a specific charged state of a representation of this algebra.
The characterization of the state by the charge is similar to the characterization of
the state by the vector $p_\mu$ as regards Lorentz transformations.
Thus one can ask if, in a similar way to what happens
with Lorentz transformations,
the unitary realization
of charge (or color) rotations is vitiated by the infrared properties of the dressing
factor. 
These are the questions of interest in this paper.
We note that the possibility of 
breaking of color rotations
has been suggested and arguments in its favor given
previously \cite{bal2}.

In a charged sector of a gauge theory like electrodynamics, we have
a collection of charged particles described by matter fields.
We can construct an action
where we separate out the coupling of the soft photons to these charged
particles.
This is explained in the next section. The action is further justified by
showing that it leads to the dressing factor and the vanishing of the overlap of
states
with different Lorentz transforms of the defining time-like vector $p_\mu$.
This is carried out in section 3.  In section 4 we give the generalization to
nonabelian gauge fields. The color rotations are constructed, and,
in section 5,
in an approximation spelled out in detail there, we show the vanishing of
the overlap of color-rotated states.
Some of the details which are technical and not essential to the flow of logic
are relegated to three appendices.

The problem of color rotations is clearly of import to the question of color confinement.
We end the paper with a short discussion of this matter. Previous work
by Balachandran and collaborators, including the discussion of the breaking
of Lorentz and color transformations, are in \cite{{bal1}, {bal2}}. The present paper has 
occasional overlap with those papers.

\section{An action for infrared dynamics}

One approach to the infrared dynamics is via the algebra of local observables and
careful analysis of the asymptotic behavior, as carried  out in many papers,
see for example \cite{{buchh}, {morchio}, {bal1}} and references therein.
Perhaps a more physical point of view is obtained by noting that, basically, 
in a charged sector of the theory we have a
collection of charged particles (described by matter fields) with a 
net charge not equal to zero.
We may think of this cluster of particles as a single composite particle.
In terms of coupling to the photon, it is well known that the low energy
photons (with $\omega \sim \vert\vk\vert \rightarrow 0$) only couple
to the overall charge, the monopole moment of the charge distribution, while
photons of higher energy couple to the higher moments of the charge distribution
via $F_{\mu\nu}$ and its derivatives. The soft photons decouple from the
higher moments as $\vk \rightarrow 0$ since they involve
$F_{\mu\nu}$ and its derivatives.
Thus it should be possible to separate out
the overall dynamics of the system of charged particles and hard photons from
the infrared
photons. This can be done as follows.

We start by considering a set of fields which are confined to a region of space
on a given time-slice; for brevity, we refer to this as a droplet of fields.
Under time-evolution, the fields will have their own dynamics, but
we will also have the overall propagation of this droplet.
The action is given by
\beq
S = {1\over 2} \int_\Omega d^4x~ \sqrt{-\eta} ~ \eta^{\mu\nu} \del_\mu \phi \, \del_\nu \phi
\label{1}
\eeq
For simplicity, we consider a single free scalar field as this suffices to illustrate the 
main points of our discussion. The metric of spacetime is denoted
by $\eta_{\mu\nu}$.
In (\ref{1}), the boundary of the region $\Omega$
can be quite complicated, because of the boundary conditions for the
fields and also because, generically, the fields inside can cause deformations
 of the surface $\del \Omega$
as well. The simplest strategy to describe the dynamics is to think of $\Omega$ as 
the embedding of a 4-volume in the spacetime. In other words, we consider a world volume
with coordinates $\xi^a$, $a= 0, 1, 2, 3,$
with prescribed boundaries for the spatial $\xi$'s. The embedding
$x^\mu (\xi)$ gives the volume traced out  by the droplet. The action, which to
begin with must be in spacetime, can be pulled back to
the world volume defined by the $\xi$'s. Thus we may write (\ref{1}) as
\beqar
S &=& {1\over 2} \int d^4\xi~ \det\left| {\del x\over \del \xi} \right|
\sqrt{-\eta} ~ \eta^{\mu\nu} {\del \xi^a \over \del x^\mu} {\del \xi^b \over \del x^\nu} 
{\del\phi\over \del \xi^a} {\del \phi \over \del \xi^b}\nonumber\\
&=&{1\over 2} \int d^4\xi \sqrt{- g}~ g^{ab} \, {\del\phi\over \del \xi^a} {\del \phi \over \del \xi^b}
\label{2}
\eeqar
where $g_{ab}$ is the induced metric given by
\beq
g_{ab} = \eta_{\mu\nu} {\del x^\mu \over \del \xi^a} {\del x^\nu \over \del \xi^b}
\label{3}
\eeq
The induced metric depends on the ambient metric $\eta_{\mu\nu}$ and the embedding.
We can think of $x^\mu (\xi )$ as additional degrees of freedom describing the
motion of the droplet. With the action (\ref{2}), we can describe the quantum
dynamics of the fields $\phi$ and $x^\mu (\xi )$ using the path-integral formalism.
Notice that (\ref{2}) is the standard action  for a scalar field,
so the variational equations or the path-integral will be the standard ones in terms of the
target space coordinates. (The target space is the real spacetime here.)
The new ingredient therefore has to do with the embedding map.
In principle, we must integrate over this in path-integral to
obtain the full quantum dynamics, but for our purpose here,
we can do a simpler 
analysis at the classical level (juts for the embedding map).
Classically the variational equation for $x^\mu (\xi )$ should describe the  motion
of the droplet. We will now show this result; our purpose is to demonstrate that the fields
in $\Omega$ behave as a single composite
particle. This result is not absolutely essential to our discussion
of electrodynamics; the main line of reasoning picks up after equation
(\ref{8}).

The variation of the action under a change of the induced metric can be written as
\beq
\delta S = - {1\over 2} \int d^4\xi~ \sqrt{-g }~ \delta g_{a b} \, T^{ab}
\label{4}
\eeq
This is the definition of the energy-momentum tensor in the world volume coordinates.
There are two related statements we can obtain from this equation.
The
action (\ref{2}) has full diffeomorphism invariance in terms of $\xi$; i.e. we could
use $\xi^{a'} = \xi^a + \epsilon^a(\xi)$ in place of $\xi^a$. For infinitesimal
$\epsilon^a$ this corresponds to the change
$ \delta g_{ab} = - g_{cb} \del_a\epsilon^c - g_{ac} \del_b \epsilon^c$.
The invariance of the action leads to the (covariant) conservation of the energy-momentum tensor,
\beq
\del_a (T^{ab} \sqrt{- g} ) = 0
\label{4a}
\eeq
We can also consider the variation of the embedding map
$x^\mu (\xi)$. In this case, we have
\beq
\delta g_{ab} = \delta x^\alpha\, {\del \eta_{\mu\nu} \over \del X^\alpha}\, \del_a x^\mu \del_b x^\nu
+ \eta_{\alpha \nu} \del_a (\delta x^\alpha )\, \del_b x^\nu + \eta_{\mu \alpha} \del_a x^\mu\, \del_b (\delta x^\alpha)
\label{4b}
\eeq
Using this in (\ref{4}) and using the conservation law (\ref{4a}) we find the equation of motion for
$x^\mu (\xi )$ as
\beq
T^{ab} \left[ {\del^2 x^\nu \over \del\xi^a \del\xi^b} + \Gamma^\nu_{\alpha\beta}
{\del x^\alpha \over \del \xi^a} {\del x^\beta \over \del \xi^b}
\right] = 0
\label{5}
\eeq
where $\Gamma^\nu_{\alpha\beta}$ is the standard Christoffel symbol for
$\eta_{\mu\nu}$.
In the quantum theory, this is truly an operator equation, but if we choose to
approximate the overall dynamics classically, but keep the full quantum dynamics for
$\phi$, this will become
\beq
\la T^{ab}\ra  \left[ {\del^2 x^\nu \over \del\xi^a \del\xi^b} + \Gamma^\nu_{\alpha\beta}
{\del x^\alpha \over \del \xi^a} {\del x^\beta \over \del \xi^b}
\right] = 0
\label{6}
\eeq
where the expectation value is in the chosen state for $\phi$ fields.

We now consider a simple example to illustrate how the dynamics works out.
For this, consider the embedding
\beq
x^\nu (\xi) = \begin{cases}
x^0  = \xi^0 \hskip .2in& \nu =0\\
x^i (\tau) + u^i (\xi) \hskip .2in&\nu =i
\end{cases}
\label{7}
\eeq
where, as usual, the superscripts $0$ and $i$ refer to time and space components.
(Here we make a special choice with the world volume time-coordinate
identified as $x^0$; we also use $\tau$ for $\xi^0$ to bring the equations
to a more familiar form.) 
The spatial coordinates have a part $x^i(\tau )$ which only depends on $\xi^0 = \tau$,
while $u^i (\xi )$ can have dependence on all $\xi^a$.
Thus the coordinates $(x^0, x^i (\tau ))$ are 
like the ``center of mass" coordinates while $u^i (\xi)$ describe the 
internal dynamics of the droplet such as compressional modes,
elastic deformations, etc.
The dynamics for such
degrees of freedom are ultimately induced by the dynamics of
the field $\phi$. This can be seen from the equations of motion for
the $u^i$-part, or more generally, from the fact that
$\phi (x) $ depends on $\xi^a$ through the embedding $x^\mu (\xi )$.
If we neglect the change in $u^i (\xi)$ (which is equivalent to neglecting
all deformations of the droplet),
$\del_a x^\alpha$ has a nonzero value only for $a = 0$. Thus (\ref{6}) becomes
\beq
\la T^{00}\ra  \left[ {d^2 x^\nu \over d\tau^2}  + \Gamma^\nu_{\alpha\beta}
{d x^\alpha \over d\tau} {d x^\beta \over d\tau}
\right] = 0
\label{8}
\eeq
If additional external potentials involving $x^\alpha$ were included, we
would have obtained
a force term on the right hand side.
Inspection of (\ref{8}) shows that the droplet propagates as
a particle of mass $\int d^3\xi\, \sqrt{- g} \la T^{00} \ra$. This shows that
the total inertia correctly includes all contributions from the quantum effects
of the $\phi$ fields as 
well.\footnote{Another reason for this example  and comment, beyond  illustrating our formulation,
 is that there was some confusion in
the recent literature on the question of whether the Casimir energy
contributes to the overall inertia of fields confined to a region \cite{casimir}.
Our analysis unambiguously shows that it will.}

We now turn to how we can apply similar reasoning to electrodynamics
in flat ambient spacetime.
For this,
we do not need the full generality of arbitrary diffeomorphisms. 
It is sufficient to take $x^\mu (\xi)$ as the Poincar\'e transform of $\xi^a$. The parameters
of the transformation will depend on $\xi^0$ (which is the time-variable before
translations).
Further, for the soft photon
absorption and emission amplitudes, the spin of the charged particle is not
important, only the translation part of the Poincar\'e transform is relevant.
Thus we
will take $x^\mu (\xi )$ to be given as
\begin{align}
x^0 &= Z^0(\xi^0), 
\hskip .3in &x^i (\xi) &= \xi^i + Z^i (\xi^0)\nonumber\\
dx^0 &= {\dot Z}^0 d\xi^0 = {\dot Z}^0 dt, 
&dx^i &= d\xi^i + {\dot Z}^i dt
\label{9}
\end{align}
The corresponding induced metric is given by
\beq
ds^2 = dZ_0^2 - (d\xi^i + dZ^i )^2
= {\dot Z}_0^2 dt^2 - (d\xi^i + {\dot Z}^i dt )^2
\label{10}
\eeq
Comparing this with
the usual splitting of the metric into the 3-metric $\sigma_{ij}$
and lapse ($\alpha$) and shift ($\beta^i$) functions,
which is
given by
\beq
ds^2 = \alpha^2 dt^2 - \sigma_{ij} \,(dx^i + \beta^i dt) (dx^j + \beta^j dt) 
\label{11}
\eeq
we see that our choice
(\ref{9}) is equivalent to $\alpha = {\dot Z}_0$, ${\dot Z}_i = \beta_i$ and
$\sigma_{ij} = \delta_{ij}$.
The metric tensor corresponding to
(\ref{10}) and its inverse are given by
\beq
g_{ab} = \left[ \begin{matrix}
{\dot Z}_0^2 - {\dot Z}_i^2 & - {\dot Z}_i \\
- {\dot Z}_i & - \delta_{ij} \\ \end{matrix} \right],
\hskip .3in
g^{ab} = {1\over {\dot Z}_0^2} \left[ \begin{matrix}
1 & - {\dot Z}_i \\
- {\dot Z}_i & - \delta_{ij}  {\dot Z}_0^2  + {\dot Z}_i {\dot Z}_j \\ \end{matrix} \right]
\label{12}
\eeq
with $\det g = - {\dot Z}_0^2$.

For electrodynamics, we are aiming to separate out the dynamics of the soft modes,
so we will use $A_\mu$ as the gauge potential for the hard photons and
$a_\mu$ for the soft photons, with the corresponding field strengths
$F_{\mu\nu} = \del_\mu A_\nu - \del_\nu A_\mu$,
$f_{\mu\nu} = \del_\mu a_\nu - \del_\nu a_\mu$.
The action is then given by
\beqar
S&=& S_1 + S_2 + S_3\nonumber\\
S_1 (\Lambda, \lambda )&=& - {1\over 4}
\int d^4\xi\, \sqrt{- g} ~ g^{ac} g^{bd} \, F_{ab} F_{cd} ~+~
S_{\rm matter} ( g)\nonumber\\
S_2 (\lambda, \mu ) &=& - {1\over 4} \int d^4y\, \sqrt{-\eta} \,\eta^{\mu\alpha} \eta^{\nu\beta}
f_{\mu\nu} f_{\alpha\beta}\nonumber\\
S_3 &=& - q \int a_\mu (Z)\, dZ^\mu
\label{13}
\eeqar
In this action (\ref{13}), instead of limiting the fields to some region $\Omega$ in spacetime,
we separate the fields in terms of the range of momenta involved.
$S_1$ consists of modes of photons with wave vectors
$\vk$, with $ \lambda < \vert \vk \vert \leq \Lambda$. Thus $\Lambda$ is an upper cutoff for the
whole theory and $\lambda$ designates the separation between the hard photons and
the soft photons. If one writes this action in terms of the target space coordinates
$x^\mu$,
then it is just the standard QED action with an upper and lower cutoff on the range of momenta.
Thus this action will lead to the usual results of QED, as it should, only qualified by
the presence of the
cutoffs. (The limit of vanishingly small
$\lambda$ may be taken at the end for various infrared finite
quantities calculated using this part of
the action.)

Further, $S_2$ is the part of the action
which describes the soft photons; it has an upper cutoff of $\lambda$ and a lower cutoff of
$\mu$; eventually we will take $\mu \rightarrow 0$. 
We consider flat ambient spacetime,
with $\eta_{\mu\nu}$ as the Minkowski metric.
The variable of integration $y^\mu$ in $S_2$ is basically the spacetime coordinate,
like $x^\mu$.
We use a different letter since, for $S_2$, we do not consider it as the result of an embedding.
$S_3$ is the coupling of the
``droplet" of hard photons and charged fields, treated as single composite
particle, to the soft photons.
We assume that there is some gauge-invariant way of introducing the
cutoffs; the details of this are not relevant for our
discussion.
Further, we will take matter fields to be massive so that there are no
additional infrared problems
due to a vanishing mass for the charged particles. 
$S_1$ is the term involving the induced metric. It has $Z$-dependence which couples it to
the soft photons. However, if we consider $S_1$ in terms of $x^\mu$, 
it is just the standard QED (with a lower cutoff) coupled to charged particles, so 
up to a diffeomorphism, the theory of hard photons and charged particles is
standard QED without infrared divergences.
(We also note that lapse and shift functions or the corresponding diffeomorphisms are
central to the analysis of the BMS symmetry and soft modes for gravitons 
\cite{{grav1}, {grav2}, {grav3}}.)

The derivation of the action for the droplet of scalar fields outlined earlier gives
the physical reasons for the choice
of the action
in (\ref{13}) for electrodynamics. But ultimately, (\ref{13}) is to be taken as the
starting postulate of our analysis. The real reason for it is that it reproduces
known results for the dressing factor for charged states
in terms of the soft photons and the spontaneous breakdown of Lorentz 
symmetry as in \cite{{buchh}, {morchio}}.
To demonstrate this and work out the
consequences of (\ref{13}),
we will analyze the dynamics in a combination of the path-integral
and Hamiltonian approaches. But, as a first step,
 it is useful to consider the classical equations of motion
for the soft photons. The variation of $x^\mu (\xi )$ or $Z^\mu$ gets contributions from
$S_1$ (via the induced metric $g_{ab}$) and $S_3$, while variations
in $a_\mu$ have contributions from $S_2$, $S_3$.
The variational equations are given by
\beqar
{1\over {\sqrt{- g}}} \del_a \left[  \sqrt{- g}\, T^{ab} \,
\eta_{\alpha\nu} \del_b x^\nu \right] &=& f_{\alpha \mu } J^\mu\nonumber\\
\del_\mu f^{\mu\nu} &=& J^\nu\label{14}\\
J^\nu (y) &=& q \int d\tau \, {d Z^\mu \over d \tau} \delta^{(4)} (y - Z(\tau ))
\nonumber
\eeqar
Here $T^{ab}$ is the energy-momentum tensor 
for the matter fields and the hard modes of the electromagnetic field.
Unlike the case of the scalar field, we have a coupling of $S_1$ to $Z^\mu$ via
$S_3$, so we do not expect conservation of $T^{ab}$. Expanding out the divergence
in the first of these equations, we find
\beq
T^{ab} \left[ {\del^2 x^\nu \over \del \xi^a \del \xi^b}\right]
+ \del_b x^\nu {1\over \sqrt{- g}} {\del \over \del \xi^a} (\sqrt{- g}~T^{ab} )
= f^{\nu\mu} J_\mu
\label{15}
\eeq
The second term is the covariant divergence of $T^{ab}$,
and since $x^\nu$ is a diffeomorphism of
$\eta^a$, for this term, we can go back to the $x$-coordinates and write
\beq
\del_b x^\nu {1\over \sqrt{- g }} {\del \over \del \xi^a} (\sqrt{- g}~T^{ab} )
= {\del \over \del x^\mu} T^{\mu\nu}
\label{16}
\eeq
Further, the second equation in (\ref{14}), which is
the equation of motion for $a_\mu$, gives
\beq
\del_\mu t^{\mu\nu} = - f^{\nu\mu} J_\mu
\label{17}
\eeq
where $t^{\mu\nu}$ is the energy-momentum tensor for the electromagnetic field.
It is defined as
$t^{\mu\nu} = f^{\mu\alpha} f_{\alpha}^{\, \nu} + {\quarter} \eta^{\mu\nu} f^2$.
Conservation of energy-momentum for the whole theory reduces to
$\del_\mu (T^{\mu\nu} + t^{\mu\nu} ) = 0$. This is obtained by considering the invariance
under coordinate transformations
for the full action in terms of $x^\mu$, $y^\mu$.
The end result is that
the equation for $x^\nu$ simplifies to
\beq
T^{ab} \left[ {\del^2 x^\nu \over \del \xi^a \del \xi^b}\right] =
T^{00} {\ddot Z^\nu} = 0
\label{18}
\eeq
Thus the current 
in (\ref{14}) is given by a constant vector
${\dot Z}^\mu \equiv p^\mu$.
If we take the limit of $\lambda$ as well as $\mu$ becoming zero,
we expect energy loss due to the radiation of soft photons to go to zero
and we will get $\del_\mu T^{\mu \nu} = 0$. This is indeed the case as shown in the
appendix.

The key result in this section is the action (\ref{13}). 
We will now use it to construct the dressing factor for charged states.

\section{The dressing factor and breaking of Lorentz symmetry}

We start with the action (\ref{13}) focusing on the dynamics of the
soft photons described 
by $S_2 + S_3$ since this is sufficient for 
deriving the dressing factor.
Consider quantizing this theory using functional integrals.
For simplicity, we use the gauge
$a_0 = 0$ and $\nabla \cdot a = 0$. States can then be represented by wave functionals
of $a_i$ (which is transverse). The transition matrix element
between states $\ket{\alpha}$ and $\ket{\beta}$ is given by
\beq
\bra{\alpha} e^{-i H t} \ket{\beta} =
\int [da] \,  \Psi^*_\alpha (a')\, e^{i S[a]) } \, \Psi_\beta (a'')
\label{19}
\eeq
where $[da]$ is the appropriate gauge-invariant measure and
the configurations which are integrated over have
$a (t) = a'' $ at $t =0$ and $a(t) = a'$ at the final value of time $t$.
$\Psi_\alpha (a') $ and $\Psi_\beta (a'')$ are the 
wave functionals for the final and initial states, respectively,
corresponding to the times $t$ and $0$. The action $S[a] = S_2 + S_3$
with the addition of the needed gauge fixing.

Consider now the action where we shift the variable as
$ a\rightarrow {\hat a} + a^{(c)}$, where
$a^{(c)}$ is a background $c$-number function.
This will be chosen to obey the equations of motion
\beq
\del_\mu (f^{(c)})^{\mu\nu}  = J^\nu
\label{20}
\eeq
The action $S_2(\lambda , \mu)  +S_3(\lambda , \mu ) $ simplifies as
\beqar
S [{\hat a} + a^{(c)} ] &=& - {1\over 4} \int {\hat f}^2
- {1\over 2}\int 
(f^{(c)})^{\mu\nu} {\hat f}_{\mu\nu} - {1\over 4} \int (f^{(c)})^2 - \int {\hat a}_\mu J^\mu
- \int (a^{(c)})^\mu J_\mu\nonumber\\
&=& - {1\over 4} \int {\hat f}^2 - \oint {\hat a}_\nu (f^{(c)})^{\mu\nu} dS_\mu
+ \int {\hat a}_\nu \bigl(  \del_\mu (f^{(c)})^{\mu\nu} - J^\nu \bigr) + \cdots
\nonumber\\
&=& - {1\over 4} \int {\hat f}^2 + \int d^3x~e^{(c)}_i {\hat a}_i \Bigr]^t_0 + \cdots
\label{21}
\eeqar
where $e^{(c)}_i $ is the electric field corresponding to $a^{(c)}_i$ and the ellipsis denotes
terms which are independent of $\hat a$ and hence constants which factor out of
the functional integral.
There are gauge fixing terms for the $\hat a$-fields which are understood as being added to
the expression shown. Using the result (\ref{21}), we can simplify
(\ref{19}) as
\beq
\bra{\alpha} e^{-i H t} \ket{\beta} =
\int [d{\hat a}] \, \Bigl[ \Psi^*_\alpha (({\hat a} + a^{(c)})' , t) \, e^{i \int e^{(c)}\cdot {\hat a}  (t)}\Bigr]\,
 e^{i S_0[{\hat a} ]) }\,
\Bigl[e^{- i \int e^{(c)}\cdot {\hat a} (0)}\Psi_\beta (({\hat a} + a^{(c)})'', 0)\Bigr]
\label{22}
\eeq
where $S_0 [{\hat a}] = - \quarter \int {\hat f}^2$. This equation shows that,
as far as the dynamics of the soft photons is concerned, we have a free Maxwell theory
(governed by $S_0[{\hat a}]$)
and hence its contribution to correlators of hard photons and matter will factor out. However,
the incoming and outgoing states must be given by
\beqar
\braket{{\hat a}  | {\tilde \beta}} &=& e^{- i \int e^{(c)}\cdot {\hat a}} \, \Psi_\beta ({\hat a} + a^{(c)})\nonumber\\
&=&e^{- i \int e^{(c)}\cdot {\hat a}} \, e^{i \int a^{(c)}\cdot {\hat e} } \, \Psi_\beta ({\hat a})
\nonumber\\
&=& V ( a^{(c)} , e^{(c)} ) ~ \Psi_\beta ({\hat a})
= \bra{a} V ( a^{(c)} , e^{(c)} )\ket{\beta}
\label{23}
\eeqar
where
\beq
V ( a^{(c)} , e^{(c)} ) = \exp \left[ - i \int \left( e^{(c)}\cdot {\hat a}
- a^{(c)}\cdot {\hat e} \right) \right]
\label{24}
\eeq
Here we have translated the wave functionals
to the operator language using $\Psi_\beta ({\hat a}) = \braket{{\hat a} |\beta}$,
using the notation $\ket{{\tilde \beta}}$ for the redefined state.
This allows us to obtain the operator expression for the
dressing factor $V ( a^{(c)} , e^{(c)} )$.
In (\ref{23}, \ref{24}),
${\hat e}_i$ is the operator
conjugate to ${\hat a}_i$, so that $\exp[ i \int a^{(c)}\cdot {\hat e} ] $
can be used to shift the field ${\hat a} + a^{(c)}$ to $\hat a$ in $\Psi$.
There can be normal ordering corrections in going from the second to last 
line in (\ref{23}); this has been absorbed into the normalization of the wave functionals.
$V ( a^{(c)} , e^{(c)} )$ gives the dressing factor for the states due to the soft photon modes.
If the initial and final states in
(\ref{19}) are the vacuum states, we can take them to be
the Fock vacua for the soft photons. But the action of 
the formally unitary operator $V ( a^{(c)} , e^{(c)} )$ converts them to
coherent states defined by the classical functions
$a^{(c)}_i$ and $e^{(c)}_i$.
The full computation of the amplitude now reduces to the computation with
the hard photons and the matter fields with the dressing factor $V$ included
 for the incoming and outgoing
states.
Since the action for the hard photons has a lower cutoff of $\lambda$, we see that
there will be no infrared divergences in the
correlations functions for the hard photons and matter
fields
as $\mu \rightarrow 0$.

We now turn to the issues with the unitary implementation of Lorentz transformations.
This result can be see by considering the
overlap of the coherent states defined above.
The operators $\hat a_i$ and $\hat e_i$ have the mode expansion
\beqar
{\hat a}_i &=& \sum_k  {1\over \sqrt{2 \omega_k \V}}
\left( c_i(k) e^{-ikx} + c^\dagger_i (k) e^{ikx} \right)\nonumber\\
{\hat e}_i &=& \sum_k  {(-i\, \omega_k) \over \sqrt{2 \omega_k \V}}
\left( c_i(k) e^{-ikx} - c^\dagger_i (k) e^{ikx} \right)
\label{25}
\eeqar
where $\omega_k = \sqrt{\vk\cdot\vk}$ and $c_i (k)$, $c_i^\dagger (k)$ are the
usual annihilation and creation operators for the photons.
(Here we have used a discrete sum over $\vk$ rather than an integral,
because this is convenient for
calculations.
We take fields in a cubical volume $\V = L^3$ with
periodic boundary conditions for the fields, so that
$\vk = 2\pi (n_1, n_2, n_3)/L$, $n_i \in {\mathbb Z}$.
Eventually, we take the limit $L \rightarrow \infty$ in the usual way.)
The summation (or integration as $\V \rightarrow \infty$) is over the range
$\mu \leq \vert\vk\vert \leq \lambda$. 
The operator $V ( a^{(c)} , e^{(c)} ) $ in (\ref{24})
can be explicitly written out in terms of these operators as
$V = e^{ \Phi}$ with
\beqar
\Phi &=& - i \int \left( e^{(c)}\cdot {\hat a}
- a^{(c)}\cdot {\hat e} \right)\nonumber\\
&=&q \sum_k  {1\over \sqrt{2 \omega_k \V} }\left[
\Omega^+_i (-k, p)\, c_i(k) - \Omega_i^- (k, p) \, c_i^\dagger (k) 
\right]
\label{26}\\
\Omega_i^{\pm}(k, p) &=&  \left( p_i - k_i {\vk\cdot\vp \over \vk^2} \right) {1\over pk \pm i \epsilon}
\nonumber
\eeqar
Here $pk = p_0 k_0 - \vp\cdot\vk$.
The dressed states are characterized by the vector
$\vp$ as it appears in $\Omega^\pm$; so we will designate the
dressed state for the Fock vacuum of soft photons as $\ket{\vp}$.
Carrying out the normal ordering for
$V$, we then get
\beqar
\ket{\vp\,} &=& \exp\left[ {-{q^2\over 2} (\Omega^+, \Omega^-)} \, {- q\sum_k {1\over \sqrt{2\omega_k \V}}\, \Omega^- \cdot c^\dagger}\right]
\ket{0}\nonumber\\
(\Omega^+, \Omega^-)&=&\sum_k {1\over 2 \omega_k \V} \,\Omega^+_i(k,p)\,  \Omega^-_i(k,p)
\nonumber\\
&=&
\int {d^3k \over(2\pi)^3} {1\over 2 \omega_k} \,
\Omega^+_i(k,p)\,  \Omega^-_i(k,p)
\label{27}
\eeqar
In the last expression, we have taken the $\V \rightarrow \infty$ limit. And,
once again, the range of integration is
restricted to
$\mu \leq \vert\vk\vert \leq \lambda$. 
It is easily checked that this state is normalized,
$\braket{\vp\, | \vp\, } = 1$.

The overlap of states for $\vp$ and $\vp{\, '}$ is given by
\beqar
\braket{\vp{\, '} |\vp} &=& \exp\left[
- {q^2\over 2} (\Omega^+_p, \Omega^-_p) -  {q^2\over 2} (\Omega^+_{p'}, \Omega^-_{p'})
+  {q^2} ( \Omega^+_{p'}, \Omega^-_p )\right]\nonumber\\
\vert\braket{\vp{\, '} |\vp} \vert^2 &=& \exp\left[ -  {q^2}( \Omega^+_p - \Omega^+_{p'}, 
\Omega^-_p - \Omega^-_{p'}) \right]
\label{28}
\eeqar
The integration over the magnitude of
$\vk$ can be carried out to write
\beqar
(\Omega^+_{p'}, \Omega^-_p) &=&
{1 \over 16 \pi^3} \log(\lambda /\mu) \, (\vv{\,'}, \vv)\nonumber\\
(\vv{\,'}, \vv)&=& \int d\Omega ~ {( v'_i - {\hat k}_i \vv{\,'}\cdot {\hat k} )\over 1- \vv{\,'}\cdot{\hat k} +
i \epsilon}\, {( v'_i - {\hat k}_i \vv \cdot {\hat k} )\over 1- \vv \cdot{\hat k} +
i \epsilon}
\label{29}
\eeqar
where $\vv = \vp/p_0$,  $\vv\,' = \vp\,' /p'_0$, and the
remaining integration is over the angular degrees of freedom in
${\hat k} = \vk/\vert \vk\vert$.
The bracket $(\vv{\,'}, \vv)$ may be viewed as an inner product for
the velocities $\vv$ and $\vv{\,'}$. It is positive semi-definite.
Thus, from (\ref{28}), we see that
\beq
\vert\braket{\vp\,' |\vp} \vert^2 =  \exp\left[ - {q^2 \over 16 \pi^3} \log(\lambda/ \mu)\,
(\vv- \vv{\,'}, \vv- \vv{\,'})\right]
\label{30}
\eeq
This vanishes as $\mu \rightarrow 0$ at fixed $\lambda$, for
any choice of $\vv \neq \vv{\,'}$.
Regarding $p'_\mu$ as obtained by a Lorentz transformation
of $p_\mu$, this overlap is equivalent to
calculating the matrix element of
the operator corresponding to
the Lorentz transformation
 for the dressed soft photon states; i.e,
 if $p'_\mu = \Lambda_\mu^{~ \nu} p_\nu$, then
 $\braket{\vp{\, '} |\vp} = \bra{\vp} U_\Lambda \ket{\vp}$.
The vanishing of this matrix element
is thus equivalent to
the statement that the Lorentz transformations cannot be unitarily implemented
on the dressed states.
It is also straightforward to show that
the matrix elements of all local operators, such as integrals of creation-annihilation 
operators smeared over non-zero momenta, will also vanish as
$\mu \rightarrow 0$.
Together, these statements
constitute the spontaneous breaking of Lorentz symmetry discussed
in \cite{{buchh}, {morchio}, {bal1}}.

The physical meaning of this breaking is also clear from our derivation.
It is simply that the total momentum
of the droplet of charged particles and hard photons is a superselected parameter,
and 
cannot be changed by any operator
action due to the soft modes. 
As we remarked earlier,
 the correlation functions of local operators
 do not manifest this breaking of Lorentz symmetry in the zero charge sector
 where $q^2 = 0$.
This sector is what is relevant for
the calculation of the S-matrix elements of QED in flat space
for practical applications, so standard experiments
 will not detect this. It can only be detected if there are
 other forces such as a gravitational field which can affect the
overall motion of the droplet, or in carefully designed experiments for 
detecting phase information \cite{sem}.

To briefly recapitulate, in this section we have, starting from the action (\ref{13}),
obtained the dressing factor and the coherent
states for soft photons; these are given in (\ref{23}, \ref{24}).
Further, by considering the overlap of such coherent states, specifically
as in (\ref{30}), we have rederived the known
result on spontaneous breaking of Lorentz symmetry in electrodynamics.

\section{Action and dressing factor for nonabelian gauge theory}

Generalizing from the electromagnetic case, the
 action for the nonabelian gauge theory, with the coupling to the lapse and 
shift functions, can be written down in a straightforward way as
\beqar
S&=& S_1 + S_2 + S_3 + S_4\nonumber\\
S_1 &=& - {1\over 4} \int d^4\xi~ \sqrt{- g}\, g^{ac} g^{bd}
\, F_{ab}^A F_{cd}^A\nonumber\\
S_2&=&  - {1\over 4} \int d^4\xi~ \sqrt{- \eta}\, \eta^{\mu\alpha}
\eta^{\nu \beta} \, f_{\mu\nu}^A f_{\alpha\beta}^A\label{31}\\
S_3&=& - \int d\tau~ Q_A \,a_\mu^A \,{\dot Z}^\mu\nonumber\\
S_4&=& i \int d\tau~ \sum_k w_k\Tr (t_k\, u^{-1} {\dot u} )
\nonumber
\eeqar
The superscripts $A$ denote the color components, corresponding to a basis
of the Lie algebra of the color group $G$ 
in which the gauge transformations take values.
$u$ is an element of $G$ and
$Q_A = 2\,\sum w_k\Tr ( t_k\,u^{-1} t_A u )$, $t_A$ being the generators of the group
in the chosen basis for the Lie algebra.
Our choice of normalization for $t_A$ is $\Tr (t_A\, t_B ) = {\half} \delta_{AB}$.
$t_k$ are the diagonal generators in the same basis.
The term $S_4$ leads to a  representation of the color group
with highest weight vector $(w_1, w_2, \cdots, w_r)$
upon quantization, $r$ being the rank of the algebra \cite{{wong}, {bal3}, {geom}}.

As in the Abelian case,
the wave vectors for the fields are restricted to be
in the range $\lambda \leq \vert \vk\vert < \Lambda$  for
$S_1$,  and $\mu \leq \vert \vk\vert < \lambda$ for
$S_2$, $S_3$.
In the Abelian case, we were able to take the limit
$\mu \rightarrow 0$, which led to the orthogonality of 
the dressed states for different values of the
time-like vector $p_\mu$ which characterizes them.
In the nonabelian case, even in perturbation theory, we have a problem
since there are many self-interactions for the gauge field.
If we want to use perturbation theory, generally because of asymptotic freedom,
$\mu$ must be larger than some scale factor
like $\Lambda_{QCD}$ which defines the theory.
However, from (\ref{30}), we see that
we need $\mu \rightarrow 0$ to obtain the orthogonality of states.
Therefore, we do not have an obvious kinematic regime
where we can use perturbation theory and
make statements regarding the orthogonality
of states due to the dressing factor, even to the lowest order in the perturbative expansion.
One option is to consider a theory with sufficient number of matter fields so
that we do not have asymptotic freedom. One can then pose the question of whether
color and Lorentz transformations can be implemented
within perturbation theory in such a model. 
We will show that color transformations cannot be unitarily implemented
in such a theory, in the sectors with nonzero color charges.
The same applies to the Lorentz transformations, but that is
essentially the same as in electrodynamics, so we focus on the
color rotations.

Returning to the action, our focus is on the low energy modes
corresponding to the action $S_2 + S_3 + S_4$.
As before, we will consider the shift
\beq
a_\mu^A = {\hat a}^A_\mu + (a^{(c)})^A_\mu
\label{32}
\eeq
In the Abelian case, $a^{(c)}_\mu$ was a $c$-number field
determined in terms of the source ${\dot Z}^\mu$. In the present case,
because the source is $Q_A {\dot Z}^\mu$, 
$(a^{(c)})^A_\mu$ will involve $Q_A$which is the color charge operator
when we quantize $u \in G$. So $(a^{(c)})^A_\mu$
will no longer be a $c$-number function.

We may think of the whole quantization procedure (for both the fields and for $g$)
 in the
Hamiltonian formalism with states described by wave functionals 
of $a_i^A$ and $g$,
\beq
\Psi_{\alpha, A} (a, g) = \braket{a, g| \alpha, r}
\label{33}
\eeq
corresponding to a state $\ket{\alpha}$ labeling momenta, polarization, etc. and color
state $\ket{r}$.
We used a functional integral approach to identify the dressing factor in the
Abelian case. However, here, because of the ghosts which may not factor
out, there is no particular advantage to the
functional integral. We will use a Hamiltonian approach
with $a_0^A = 0$. (The previous results will also be easily recovered within this framework.)
The Hamiltonian for $S_2+ S_3 +S_4$ is given by
\beqar
H &=& {1\over 2} \int (e^2 + b^2 ) + a_i^A Q_A {\dot Z}^i\nonumber\\
e^A_i &=& {\dot a}^A_i, \hskip .2in
b^A_i = {1\over 2} \epsilon_{ijk} (\del_j a^A_k -
\del_k a^A_j + f^{ABC} a^B_j a^C_k )
\label{34}
\eeqar
The time-evolution operator is obviously given by
$U (t,0) = T \exp\left(  -i \int^t_0 dt\, H\right)$.
We now write this as $U(t,0) = V(t)^{-1} {\tilde U}(t,0) \, V(0)$
where
\beqar
{\tilde U} (t,0) &=& T  \exp\left(  -i \int^t_0 dt\, {\tilde H}\right)\nonumber\\
{\tilde H}&=& V \, H\, V^{-1} + i {\del V\over \del t} \, V^{-1}
\label{35}
\eeqar
This factorization, so far, is just a mathematical identity. The idea now is to choose 
$V$ such that the interaction terms are eliminated in
${\tilde H}$.
For the Abelian case, we can take the ansatz
\beq
V = \exp\left[ -i \int \left( e^{(c)}_i {\hat a}_i - a^{(c)}_i {\hat e}_i\right) \right]
\label{36}
\eeq
This leads to
\beq
{\tilde H}= {1\over 2} \int ({\hat e}^2 + {\hat b}^2 )
+ \int \left[ e^{(c)}_i {\hat e}_i + b^{(c)}_i {\hat b}_i + a^{(c)}_i J_i
+ {\hat a}_i J_i - {\dot a}^{(c)}_i {\hat e}_i + {\dot e}^{(c)}_i {\hat a}_i
\right] + c\mhyphen{\rm number~terms}
\label{37}
\eeq
where $b_i = {\half} \epsilon_{ijk} \del_j a_k$ is the magnetic field.
We can write $\int b^{(c)}_i {\hat b}_i  = - \int  \nabla^2 a^{(c)}_i \, {\hat a}_i$ via
a partial integration. We then see
that all the mixing terms can be eliminated by 
setting the coefficients of ${\hat e}_i$ and ${\hat a}_i$ in the bracketed
terms to zero, i.e., by choosing
$a^{(c)}_i$ and $e^{(c)}_i$ to be be solutions of
\beq
e^{(c)}_i - {\dot a}^{(c)}_i = 0, \hskip .2in
-\nabla^2 a^{(c)}_i + {\dot e}^{(c)}_i + J_i = 0
\label{38}
\eeq
These are the same equations as we had before for the background field.
The splitting of $U$ as
$V(t)^{-1} {\tilde U}(t,0) \, V(0)$
shows that
$V$ is the dressing factor, since the time-evolution given by
$\ket{\psi(t)} = U(t, 0) \, \ket{\psi(0)}$ translates as
\beq
\left[ V(t) \ket{\psi(t)}\right]  = {\tilde U}(t, 0) \, \left[ V(0) \ket{\psi(0)}\right]
\label{39}
\eeq
Thus the dressed states $V\ket{\psi}$ evolve with the Hamiltonian
${\tilde H}$ in which the terms mixing
$({\hat a}_i, {\hat e}_i)$ and 
$(a^{(c)}_i, e^{(c)}_i )$ have been eliminated
via (\ref{38}).
The remaining Hamiltonian shows that
the soft photons behave as a decoupled free system.
(The $c$-number terms are irrelevant for this discussion.)

A similar analysis can be done for the nonabelian case, but it is trickier for two reasons:
1) There are gluon-gluon type interactions due to the 
cubic and quartic terms in the action or Hamiltonian. 2)
The current $J^A_i = {\dot Z}_i Q^A$ is an operator (because of the $Q^A$)
and hence the analogue of the equations
(\ref{38}) will not yield
$c$-number functions for $a^{(c)}_i$, $e^{(c)}_i$.
We can parametrize $V$ in the form
\beq
V = \exp\left[ -i \int \left( e^{A(c)}_i {\hat a}^A_i - a^{A(c)}_i {\hat e}^A_i\right)  + i \, K \right]
\label{40}
\eeq
${\tilde H}$, defined as in (\ref{35}),  will now contain many additional terms compared to
(\ref{37}) consisting of the gluon-gluon interaction terms,
commutators between these and the exponent in $V$, and commutators of
$Q_A$ with various terms including $a^{A(c)}_i$ and $e^{A(c)}_i$. 
A systematic perturbation series for $K$ in
(\ref{40}) can be written down although
the calculation of the terms in the series can reveal infrared divergences.
In the spirit of defining colored states as in perturbative QCD, if we neglect
these additional corrections, the solutions will be similar to the Abelian case.
Thus, to the lowest order,
\beq
{\hat a}^A_i = \sum_k  {1\over \sqrt{2 \omega_k \V}}
\left( c^A_i(k) e^{-ikx} + c^{A\dagger}_i (k) e^{ikx} \right)
\label{40a}
\eeq
We can then write the dressing factor in (\ref{40}) as
$V = e^{\Phi}$, $\Phi = i {\hat\chi}^A Q_A$ with
\beq
{\hat\chi}^A = -i \sum_k  {1\over \sqrt{2 \omega_k \V } }\left[
\Omega^+_i (-k, p)\, c_{i}^A(k) - \Omega_i^{-}(k, p) \, c_{i}^{A\dagger} (k)
\right]
\label{41}
\eeq

In the fully interacting case, the expression for $V$ will be substantially more complicated,
but we can still make the following general argument. $V$ depends on the
canonical variables ${\hat e} ^A_i$, ${\hat a}^A_i$ and on $Q_A$; the latter dependence is
entirely from  $(e^{(c)})^A_i$ and 
$(a^{(c)})^A_i$. All the variables appear in combinations such that all color indices are
contracted to form invariants. As a result, we have the following
property:  A color rotation of $Q_A$ can be compensated  by a color rotation of
the operators ${\hat e} ^A_i$, ${\hat a}^A_i$.
Further, expectation values are  evaluated by using commutation rules
for ${\hat e} ^A_i$, ${\hat a}^A_i$ which are invariant under color rotations.
Writing $Q'_A = \D_{AB} (h)  \, Q_B$, where
$\D_{AB}(h)$ is the adjoint representation of a constant $x$-independent $h \in G$,
these features tell us that
\beq
V_p (Q'_A, {\hat e}^A_i, {\hat a}^A_i ) 
= V_p (Q_A, {\hat e}^{'A}_i, {\hat a}^{'A}_i )
\label{42}
\eeq
Here we have also indicated, via the subscript,
 the dependence of $V$ on the
time-like vector $p_\mu$. 

Consider now the overlap of two states given by
\beq
I_{rs} \equiv \bra{0,r} V^\dagger_p (Q_A, {\hat e}^A_i, {\hat a}^A_i ) 
\,V_{p'} (Q_A, {\hat e}^A_i, {\hat a}^A_i )  \ket{0,s}
\label{43}
\eeq
We insert $h \, h^{-1}$ between $V^\dagger $ and $V$ to write this as
\beqar
I_{rs} &=& \bra{0,r} V^\dagger_p (Q_A, {\hat e}^A_i, {\hat a}^A_i ) \,
h\, h^{-1}\,
V_{p'} (Q_A, {\hat e}^A_i, {\hat a}^A_i )  \ket{0,s}\nonumber\\
&=& \bra{0,r} h \, \bigl[h^{-1} V^\dagger_p (Q_A, {\hat e}^A_i, {\hat a}^A_i ) \,
h \bigr] \, \bigl[h^{-1}\,
V_{p'} (Q_A, {\hat e}^A_i, {\hat a}^A_i ) h \bigr]\, h^{-1} \ket{0,s}\nonumber\\
&=&\bra{0,r} h \, V^\dagger_p (Q'_A, {\hat e}^A_i, {\hat a}^A_i ) \, 
V_{p'} (Q'_A, {\hat e}^A_i, {\hat a}^A_i ) \, h^{-1} \ket{0,s}\nonumber\\
&=& h_{r r'} h_{ss'}\,
\bra{0,r'} V^\dagger_p (Q'_A, {\hat e}^A_i, {\hat a}^A_i ) \, 
V_{p'} (Q'_A, {\hat e}^A_i, {\hat a}^A_i ) \ket{0,s'}\nonumber\\
&=&  h_{r r'} h_{ss'}\,
\bra{0,r'} V^\dagger_p (Q_A, {\hat e}^A_i, {\hat a}^A_i ) \, 
V_{p'} (Q_A, {\hat e}^A_i, {\hat a}^A_i ) \ket{0,s'}\nonumber\\
&=& h_{r r'} h_{ss'}\, I_{r's'}
\label{44}
\eeqar
This shows that $I_{rs}$ is an invariant tensor in the representation of the color charge
algebra. As a result, we should have
$I_{rs} = I\, \delta_{rs}$. This argument shows that the
 state of the color charge
is not changed by the dressing factor.
Time-evolution involved
is given by the Hamiltonian ${\tilde H}$ which is invariant under color rotations
and hence time-evolution also will not change the state of the color charge.
This is a reflection of the superselection of color charge.

This result by itself is not sufficient to
make any definite conclusion about the 
unitary implementation of color 
transformations. It is needed but is only part of the
argument.
We will now argue that any attempt to carry out a color
rotation on the coherent states will run into problems.
For this, consider the matrix element
of a color rotation operator
$h(\theta)  = \exp (i \theta^A Q_A )$ between arbitrary dressed states given by
\beq
M_{rs} = \bra{0,r} V^\dagger (Q, {\hat e}, {\hat a}) \, h (\theta) \, 
V(Q, {\hat e}, {\hat a})\ket{0,s}
\label{45a}
\eeq
We will show, albeit to lowest order in perturbation
theory (i.e., in the limit of neglecting
self-interactions of the gluon),
 that this matrix element vanishes for any $h\neq 1$ as
 $\mu \rightarrow 0$,
in a way similar to what happens for Lorentz transformations
of $p_\mu$. 
This means that color transformations cannot be unitarily implemented on the
dressed states  in the sector with net nonzero color.
A dressed state like
$V(Q, {\hat e}, {\hat a})\ket{0,s}$ where $\ket{0,s}$ is the Fock vacuum for the
infrared gluons, but can have charged particles characterized by the color index
$s$
are precisely what we need as the
charged states to be used in the lowest order in perturbation theory.
So the implication of our result is that
color transformations cannot be unitarily realized
on the perturbative states.

To summarize, we have the nonabelian generalization of the action (\ref{13}) given 
in (\ref{31}). The dressing factor, in the approximation of neglecting gluon-gluon
interactions was obtained. We also obtained superselection of color charge states;
this was obtained before in another way in \cite{bal2}.
And finally, we claim that the matrix element for a nontrivial color
rotation on the coherent states is zero. The argument
for the last statement, because it involves many technical points, is 
given below as a separate section.

\section{Vanishing matrix elements for color rotation}

Consider the matrix element of the color rotation in (\ref{45a}).
Using $1 = h h^{-1} $ to the left of $V^\dagger$, where $h$ is an element of the color group,
 we can write it as
\beqar
M_{rs}
&=& \bra{0,r} h(\theta) \, \bigl[ h^{-1}(\theta) V^\dagger (Q, {\hat e}, {\hat a}) \, h (\theta) \bigr]
\,  V(Q, {\hat e}, {\hat a})\ket{0,s}\nonumber\\
&=& h_{rq} \bra{0,q}V^\dagger (Q', {\hat e}, {\hat a}) \,
V(Q, {\hat e}, {\hat a})\ket{0,s}
\label{45b}
\eeqar
where $Q'_A = h^{-1} Q_A h = \D_{AB}(h) \,Q_B$. 
We have chosen different color states $\ket{r}$, $\ket{s}$ for $M_{rs}$.
Since 
\beq
V^\dagger (Q')  V(Q)\, [V^\dagger (Q')  V(Q)]^\dagger = 1,
\label{46}
\eeq
$V^\dagger (Q')  V(Q)$ is unitary and
we see, by the Cauchy-Schwarz inequality that the off-diagonal matrix element
$ \bra{0,q}V^\dagger (Q')  V(Q)\ket{0,s}$ does not exceed in magnitude the diagonal
element
$ \bra{0,r}V^\dagger (Q')  V(Q)\ket{0,r}$.
The vanishing of the latter is thus sufficient to show
the vanishing
of color rotation matrix element in (\ref{45a}, \ref{45b}).
So our aim is to show that the diagonal element
$ \bra{0,r}V^\dagger (Q')  V(Q)\ket{0,r}$ vanishes, for $Q' \neq Q$, as $\mu \rightarrow 0$,
 in the limit of neglecting gluon-gluon 
 interactions.\footnote{For a complete proof of color breaking, we must also show that the above matrix element vanishes when $V(Q)$ is replaced by $\alpha V(Q)$ where $\alpha$ is any local observable . We will omit this proof here. An alternative proof of the result of this section is in \cite{bal2}.}

However establishing the vanishing of $ \bra{0,r}V^\dagger (Q')  V(Q)\ket{0,r}$ is
not so straightforward because, even if we pick $Q$'s in an Abelian
subalgebra, $Q'_A$ can have components outside of this subalgebra
 leading to
$[Q'_A, Q_B ] \neq 0$.
Secondly, even though we can write $V = e^\Phi$ with $\Phi = i \chi^A Q_A$ given as in
(\ref{41}), the normal ordering of this expression is also not straightforward
since
\beq
[ Q_A \,c^A_i, Q_B \,c^{B\dagger}_j ] = Q^2 \delta_{ij} + Q_A\, c^{B\dagger}_j
(-i f^{ABC} ) \, c^C_i
\label{47}
\eeq
and the second term on the right hand side can generate further commutator terms
in reordering the series for $V$.

So, towards carrying out
 the calculation of the matrix element $ \bra{0,r}V^\dagger (Q')  V(Q)\ket{0,r}$, we first consider
\beq
\bra{0,r} V(Q) \ket{0,r} = \bra{0,r} e^{i {\hat\chi}^A Q_A}  \ket{0,r}
\equiv e^{- {\half } W}
\label{48}
\eeq
This is the definition of $W$. By the general properties of such
expectation values,
we can write $W$ in terms of connected functions of ${\hat\chi}^A Q_A$. For this we note that
${\hat\chi}^A$ is a free bosonic field, although it has many components,
so it has only connected two-point functions given by
\beq
\la {\hat\chi}^A \, {\hat\chi}^B \ra = z\, \delta^{AB} 
\label{49}
\eeq
where we use 
\beq
z = ( \Omega^+, \Omega^-) = {1\over 16\pi^3 } \log(\lambda/\mu) \, (\vv, \vv )
\label{50}
\eeq
This variable $z$ takes real values from zero to infinity, with
$z \rightarrow \infty$ corresponding to $\mu \rightarrow 0$.
Equation (\ref{49}) shows that we can evaluate $W$ in terms of a series of Wick contractions
on powers of $\chi$.
It is easy to see that $W$ is real, with $W \geq 0$. It is also of the form
\beq
W = 2 \,Q^2  z \sum_0^\infty w_n \,\left( {z\,C_{\rm ad}  \over 2}\right)^n
\label{51}
\eeq
where $w_n$ are numerical coefficients and $C_{\rm ad}$ is the quadratic
Casimir for the adjoint representation of the 
group. We show these properties in the appendix where we also find
\beq
w_0 = {1\over 2} , \hskip .2in
w_1 =  {1\over 4!} , \hskip .2in
w_2 = {5 \over 6!}
\label{52}
\eeq

Another property of importance to us
is that, viewing $W$ as a function of $z$, it obeys the 
inequality
\beq
W(z) \geq W(z_1) , \hskip .3in z\geq z_1
\label{53}
\eeq
The reasoning behind this result is that, for $z > z_1$, more states are included 
because we are increasing the range of $\vk$-values and hence we should
expect some such condition following from the
completeness relation.
More specifically, the required expectation value can be written as
\beq
\la e^\Phi \ra = \la e^{ i {\hat\chi}\cdot Q}\ra = \la e^{-i {\hat\chi}\cdot Q}\ra
= \la \cos {\hat\chi}\cdot Q\ra = \la 1 - 2 \, \sin^2 ({\hat\chi}\cdot Q /2)\ra
\label{54}
\eeq
Here we have used the fact that only even powers of ${\hat\chi}$ can contribute and that
the trigonometric
relations hold as operator statements
since each expression is defined by the power series
expansion. From (\ref{54}),
$2 \, \la \sin^2 \left({\hat\chi}\cdot Q/ 2\right)\ra = 1 - e^{- W/2}$. 
We now have
\beqar
\la \sin^2 \left({{\hat\chi}\cdot Q\over 2}\right)\ra_z &=&
\sum_\alpha \bra{0,r} \sin\left( {{\hat\chi}\cdot Q\over 2}\right) \ket{\alpha, s}\,
 \bra{\alpha, s}
\sin\left({{\hat\chi}\cdot Q\over 2}\right)\ket{0,r}\nonumber\\
&=&\la \sin^2 \left({{\hat\chi}\cdot Q\over 2}\right)\ra_{z_1}
+ \sum_{\alpha'} 
\big\vert\bra{0,r} \sin\left({{\hat\chi}\cdot Q\over 2}\right)\ket{\alpha', s}\big\vert^2\nonumber\\
&\leq &\la \sin^2 \left({{\hat\chi}\cdot Q\over 2}\right)\ra_{z_1}
\label{55}
\eeqar
In this equation $\alpha'$ indicates states which involve at least one value of
$\vk$ beyond the range given by $z_1$.
Using the inequality (\ref{55}) in the expression for $W$, we find the result
(\ref{53}).

The property (\ref{53}) of $W$ shows that its derivative with respect to
$z$ is non-negative. 
(In fact, the derivative will be positive since
we obtain $W(z) = W(z_1)$ only if there are no states
with $\vk $ beyond the range given by $z_1$.)
The next property we need is that 
$W(z)$ does not saturate to a finite value as $z\rightarrow \infty$. In fact
\beq
W(z) \rightarrow \infty \hskip .2in {\rm as} \hskip .2in
z\rightarrow \infty
\label{56}
\eeq
For showing this result, we express $W$ as an integral.
As mentioned earlier,  ${\hat\chi}^A$ is a free bosonic field, and hence
the only connected correlators are the two-point ones. Therefore we can represent it
using a Gaussian integral, i.e.,
\beq
\bra{0, r} e^{i {\hat\chi} \cdot Q} \ket{0, r}
=  {1\over (2 \pi z)^{{\rm dim}G/2}} \int  [d\sigma] \, \exp\left( - {1\over 2 z} \sigma^A \sigma^A \right)
~ \bra{r} e^{i \sigma\cdot Q} \ket{r}
\label{57}
\eeq
(The $\sigma$'s are $c$-number variables.)
This is a finite dimensional integral, there are ${\rm dim}G$ independent variables
$\chi^A$ which are real. The quantity $\bra{r} e^{i \sigma\cdot Q} \ket{r}$
in the integrand
is a function of the $\sigma$'s; it is the diagonal matrix element
$g_{rr}$ of the group element
$g = \exp( i \sigma\cdot Q)$ in the representation corresponding to
the states $\ket{r}$.
However, it should be kept in mind that the integration is not over the group
volume (i.e., not with the Haar measure), but over all real values of the group parameters, $\sigma^A$ in this case.
Now $\bra{r} e^{i \sigma \cdot Q}\ket{r}$ must be a periodic function of the $\sigma$'s and
so it can be expanded in a Fourier series as
\beq
\bra{r} e^{i \sigma \cdot Q}\ket{r} = \sum_{\{n\}} C_{n_1, n_2, \cdots, n_d}
\sin\left( {n_1 \sigma_1 \over l_1}\right)\cdots
\sin\left( {n_d \sigma_d \over l_d } \right)
+ \cdots
\label{58}
\eeq
where $d = {\rm dim} G$ and the ellipsis indicates 
terms with cosines and mixtures of sines and cosines.
We may not have the same period for all 
$\sigma$'s; that will depend on the normalization of the generators, so we include
parameters $l_1, \cdots, l_d$. The key point is that, with the Gaussian measure
(\ref{57}), the average of any periodic function goes to zero
as $z \rightarrow \infty$ as seen from
\beqar
\la \sin \left( {n_1 \sigma_1 \over l_1}\right)\ra
&=& {1\over 2 i} \left[ \exp\left( - n_1^2 z /l_1^2\right) -
{\rm complex ~conjugate}\right]
\xrightarrow[z \to \infty] {} 0\nonumber\\
\la \cos \left( {n_1 \sigma_1 \over l_1}\right)\ra
&=& {1\over 2 }\, \left[ \exp\left( - n_1^2 z /l_1^2\right) +
{\rm complex ~conjugate}\right]
\xrightarrow[z \to \infty] {} 0
\label{59}
\eeqar
This is equivalent to the statement in (\ref{56}).

Finally we come to the matrix element of interest, namely,
$ \bra{0,r}V^\dagger (Q')  V(Q)\ket{0,r}$, with $Q' = h^{-1} Q h\neq Q$.
From the Cauchy-Schwarz inequality, this obeys
\beq
\vert\bra{0,r}V^\dagger (Q')  V(Q)\ket{0,r}\vert^2
\leq \bra{0,r}V^\dagger (Q')  V(Q')\ket{0,r}
\bra{0,r}V^\dagger (Q)  V(Q)\ket{0,r}
\label{60a}
\eeq
Since the right hand side is $1$, we have
\beq
\vert\bra{0,r}V^\dagger (Q')  V(Q)\ket{0,r}\vert^2
\leq 1
\label{60b}
\eeq
Equality is obtained only for $Q' =  Q$, i.e., for
$h = 1$.
We can now write this matrix element also as an integral over $\sigma$'s,
\beq
\bra{0, r} e^{- i {\hat\chi} \cdot Q'}  e^{i {\hat\chi} \cdot Q} \ket{0, r}
=  {1\over (2 \pi z)^{{\rm dim}G/2}} \int  [d\sigma] \, \exp\left( - {1\over 2 z} \sigma^A \sigma^A \right)
~ \bra{r} e^{- i \sigma \cdot Q'}  \,e^{i \sigma \cdot Q} \ket{r}
\label{61}
\eeq
Using $Q'_A = h^{-1} Q_A h$,
\beq
 \bra{r} e^{- i \sigma \cdot Q'}  \,e^{i \sigma\cdot Q} \ket{r}
 = h^{-1}_{rs} h_{pq} 
  \bra{s} e^{- i \sigma \cdot Q} \ket{p} \, \bra{q} e^{ i \sigma \cdot Q} \ket{r}
  \label{62}
  \eeq
  Once again, the product $  \bra{s} e^{- i \sigma \cdot Q} \ket{p} \, \bra{q} e^{ i \sigma \cdot Q} \ket{r}$
  is a periodic function of the $\chi$'s, not equal to $1$ for all $\sigma$.
  In fact, it is easy to see that this is not identically $1$ by 
  considering $h$ near the identity. Being periodic, 
 it can be expanded in a Fourier series and by the result
  (\ref{59}), the average matrix element in
  (\ref{61}) will give zero.
  This proves the vanishing of the matrix element
  $\bra{0, r} e^{- i {\hat\chi} \cdot Q'}  e^{i {\hat\chi} \cdot Q} \ket{0, r}$.
  In turn, this
  shows that the matrix elements of the color rotation
  in (\ref{45b}) will vanish for any $h \neq 1$, as $z\rightarrow \infty$,
  or as $\mu \rightarrow 0$.
And finally, this leads to the statement given after (\ref{45a}) about the problem with a unitary
 realization of
color transformations in the sector with net nonzero color.

\section{The problem of defining color confinement}

In this paper, we have analyzed some of the infrared issues in a
gauge theory in terms of a particular splitting of the action
into a part for the hard modes and a part for the soft modes, with 
a diffeomorphism for the hard part. This diffeomorphism which can be
viewed in terms of the lapse and shift functions for the metric encodes
the overall motion of the collection of charged particles. These functions also characterize
the time-like vector which defines the dressing of the Fock states for the
soft fields and hence the coherent states of soft modes to be used
in evaluating amplitudes.
Our results reproduce in a novel way the known results regarding the 
spontaneous breaking of Lorentz symmetry in the charged  sectors of
quantum electrodynamics.

We have also analyzed the question of a spontaneous breaking of color
transformations in a nonabelian gauge theory.
This analysis can be done explicitly if one neglects gluon-gluon interactions.
Theories where such an approximation can be valid would include
those with sufficient number of matter fields so that asymptotic freedom
is not obtained or those which admit a nonabelian
Coulomb phase.
Our results do not directly impact standard perturbative QCD
since we do have asymptotic freedom in this case.
Also, for QCD calculations, one can restrict oneself to
states with total charge equal to zero, i.e., to states which are QCD singlets.
But our results
do show
that perturbation theory with arbitrarily low momenta for the gluons
is not consistent,
since gluon single particle states would not allow
the unitary implementation of color transformations.
Our analysis also shows that
defining asymptotically nonfree nonabelian theories or nonabelian
 Coulomb phases can be problematic.
 This also calls into question the definition of color confinement.
 The standard lore has been that we should implement the Gauss law
 with test functions which vanish at spatial infinity 
 on physical states. This would still allow for charged states
 which transform as some representation of the color group.
 Thus, {\it a priori} there is no obstruction to defining color transformations
 on the single particle states.
 Confinement is then the statement that, {\it for dynamical reasons},
 the spectrum of the Hamiltonian does not contain the charged states.
 One can use, in some cases where there are no fields
transforming according to the fundamental
 representation, the expectation value of the Wilson loop
 as a diagnostic for confinement.
 This is the more conventional signature one looks for in, say, lattice simulations.
 However, our results go further.
 If, as we have argued,
 color transformations cannot be defined on perturbative gluon
 states, then we should expect that either
there cannot be gluon states of arbitrarily low energy (so that the theory can
evade our analysis by not permitting us to take $z \rightarrow \infty$)
or that single charged excitations carry infinite energy.
One can have configurations of zero total charge without
the cluster property so that single charged states cannot be separated off.
These argument clearly indicate
that our analysis has implications for the mass gap and confinement
issues in the nonabelian theory. Further analysis is needed to clarify the situation.

\vskip .15in
This research was supported in part by the U.S.\ National Science
Foundation grant PHY-1519449
and by PSC-CUNY awards.

\section*{Appendix A: Remarks on the canonical framework for $Z^\mu$ in (\ref{13})}
\def\theequation{A\arabic{equation}}
\setcounter{equation}{0}

In this appendix, we will work out some general features of the action
(\ref{13}). We will also add a Dirac action as the matter part.
The induced metric has frame fields given by
$\E^0 = {\dot Z}_0 dt$, $\E^i = d\xi^i + {\dot Z}^i dt$.
These obey $d\E^a = 0$, so that the spin connection may be taken as zero.
Thus the Dirac action has the form
\beqar
S_{\rm Dirac} &=&\int d^4\xi \sqrt{- g}~ {\bar \psi} \left[ i (\E^{-1})^\mu_a \gamma^a 
D_\mu - m \right] \psi \nonumber\\
&=& \int d^4\xi\,  {\dot Z}_0 ~
\, {\bar \psi} \left[ i {\gamma^0 \over {\dot Z}_0} (D_0 - {\dot Z}^i D_i )
+ i \gamma^i D_i - m \right] \psi
\label{A1}
\eeqar
The electromagnetic part of the action $S_1$ simplifies as
\beqar
S_{1{\rm em}} &=& - {1\over 4} \int d^4\xi \left[ - 2 {E^2 \over {\dot Z}_0}
+ 2\, {\dot Z}_0 B^2
+ 4 {{\dot Z}_i ({\vec E} \times {\vec B})_i \over {\dot Z}_0}
- 2 {\left( {\dot Z}_i {\dot Z}_i B^2 - ({\dot Z}_i B_i )^2 \right) \over {\dot Z}_0}\right]
\label{A2}
\eeqar
Notice that $v_i \equiv {\dot Z}_i / {\dot Z}_0$ is independent of the spatial
coordinates, so that we have $\int d^3\xi\, {{\dot Z}_i ({\vec E} \times {\vec B})_i / {\dot Z}_0}
=  {{\dot Z}_i / {\dot Z}_0} \int d^3\xi\,({\vec E} \times {\vec B})_i = 0$, by virtue of 
rotational invariance.
This will remove one of the terms in (\ref{A2}).
Further, 
we have invariance under reparametrization of the variable $t$;
The action can be written entirely in terms of 
$Z_0$ and the three spatial coordinates using
$d^4x {\dot Z}_0 = d^3x dZ_0$, etc.
This invariance will lead to
a zero Hamiltonian for the evolution with respect to $t$. The evolution with respect to
$x^0$ is what is relevant; equivalently, we can also make the gauge choice
${\dot Z}_0 = 1$. In this case, 
\beqar
S_{1{\rm em}} &=& \int d^4\xi\, \left[ {1\over 2} (E^2 - B^2 ) + {1\over 2} (v^2 B^2 - \vv\cdot {\vec B})^2
\right]\nonumber\\
S_{\rm Dirac} &=& \int d^4\xi \, {\bar \psi} \left[
i  \gamma^0 \del_0  - i \gamma^0 \vv\cdot {\vec D} + i \gamma^i D_i - m \right]\psi
\label{A3}
\eeqar
For the infrared modes, we have
\beq
S_2 + S_3 = {1\over 2} \int (e^2 - b^2) - q \int a_i v^i
\label{A4}
\eeq
where $e_i = \del_0 a_i$, $b_i = (\nabla \times a)_i$. We have also chosen the 
$A_0 = a_0 = 0$ gauge. 
The Hamiltonian can now be worked out as
\beq
H = \int \left[
{1\over 2} (E^2 + B^2) + {1\over 2} \left( v^2 B^2 - (\vv\cdot {\vec B})^2 \right)
- {\bar \psi} ( i \gamma^i D_i - m ) \psi + {1\over 2} (e^2 + b^2 )\right]
\label{A5}
\eeq
This is manifestly positive, except for the Dirac term, which, as usual, is not positive.
Upon quantization, the Dirac Hamiltonian will also be positive
after the usual redefinition of the negative energy states.

It is also useful to look at the dynamics of $Z^i$. Assuming that $\int B^2$ is small, or
that we can neglect it as a first approximation, we see that
\beq
{\del L \over \del v^i} = \int \psi^\dagger (-i D_i \psi ) - q\, a_i (Z)
\equiv P_i
\label{A6}
\eeq
Notice that we cannot solve this for $v^i$ in terms of the momentum $P_i$. 
From the equation of motion for $Z^i$, we get
\beq
{d \over d t}  \int \psi^\dagger (-i D_i \psi ) = q ( e_i + f_{ji} {\dot Z}^j )
\label{A7}
\eeq
As we show in appendix B, the right hand side of
this equation vanishes, showing that $\int \psi^\dagger (-i D_i \psi )$
is independent of time. This is the total momentum of the particle system.
(If we had kept the contribution for the $B_i (v^2 \delta^{ij} - v^i v^j ) B_j$-term
in the action, there would be some correction to the total momentum; qualitative
features will not be changed.)
To relate $P_i$ to $v^i$, we have to use the Hamiltonian.
Since $P_i$ is conjugate to $Z^i$, we get
\beq
[ Z^i, P_j ] = i \delta^i_j, \hskip .3in
i {\dot Z}^i = [ Z^i, H]
\label{A8}
\eeq
Thus if the Hamiltonian is expressed using $P_i$ (among other variables)
for some charged particle state, then (\ref{A8}) can relate 
${\dot Z}^i$ and $P_i$.

\section*{Appendix B: The classical infrared fields}
\def\theequation{B\arabic{equation}}
\setcounter{equation}{0}

We start by calculating the solution of the equation of motion
for $a_\mu$ which is needed to evaluate
the factor $f^{\mu\nu}J_\nu$ used in section 3.
 This solution is also what is designated as
$a^{(c)}_\mu$ in section 4. The source current is given by
\beqar
J_\mu &=& q \int d\tau\, {dZ_\mu \over d\tau} \, \delta^{(4)}(y - Z(\tau))
\nonumber\\
&=& q \int d\tau\, p_\mu \, \delta^{(4)}(y - Z(0) - p\, \tau)
\label{B1}
\eeqar
where $p_\mu = {dZ_\mu /d \tau}$. The solution in
the gauge  $a_0 =0$, $\nabla\cdot a = 0$  is given by
\beqar
a^{(c)}_i &=& - q \int d\tau {d^4k \over (2\pi)^4} 
\left( p_i - k_i {\vk\cdot\vp \over \vk^2} \right) \, {1\over k^2 + i\epsilon}
e^{-ik (x - Z(0) - p \tau )}\nonumber\\
e^{(c)}_i&=& q \int d\tau {d^4k \over (2\pi)^4} 
\left( p_i - k_i {\vk\cdot\vp \over \vk^2} \right) \, {i k_0\over k^2 + i\epsilon}
e^{-ik (x - Z(0) - p \tau )}
\label{B2}
\eeqar
We have used the time-ordered (or Feynman) Green's function as this will be
what is relevant in
the functional integral in section 4.
The electric field is given as $e^{(c)}_i = {\dot a^{(c)}}_i$. In $f^{\mu\nu} J_\nu$, first consider the
term 
\beq
f^{0i} J_i = e^{(c)}_i J_i = \int d \tau ~e^{(c)}_i (Z)\, p_i ~\delta^{(4)}( x - Z(\tau ) )
\label{B3}
\eeq
$Z^\mu(\tau)$ is the world line corresponding to the current or the overall
motion of the droplet.
In (\ref{B3}), the field $e^{(c)}_i$ is evaluated on this world line.
Taking account of the $i \epsilon$ prescription,
\beq
e^{(c)}(Z)_i = {q \over 2} \int {d^3 k \over (2\pi)^3} \,
\left( p_i - k_i {\vk\cdot\vp \over \vk^2} \right)\, \Bigl[
\int_{-\infty}^\tau d\tau' e^{-ikp (\tau -\tau')} - \int_\tau^\infty d\tau' e^{ikp(\tau -\tau')}
\Bigr]
\label{B4}
\eeq
In the first of the $\tau$-integrals, we change variables to
$u = \tau' -\tau$ and in the second to
$u = \tau -\tau'$ to get
\beq
\int_{-\infty}^\tau d\tau' e^{-ikp (\tau -\tau')} - \int_\tau^\infty d\tau' e^{ikp(\tau -\tau')}
= \int_{-\infty}^0 du \, e^{ikp u} - \int_{-\infty}^0  du\, e^{ikp u}  = 0
\label{B5}
\eeq
This shows that the electric field evaluated on the world line itself is zero. Consider now the magnetic field which
contributes to $f^{ij} J_j$. Evaluating it on the world line
we find
\beq
\epsilon_{iab} \del_a a^{(c)}_b =
(-i q) \epsilon_{iab} \, p_b \int {d\tau'}{d^4k \over (2\pi)^4} {k_a\over k^2 + i \epsilon}
e^{-ikp(\tau -\tau')}
\label{B6}
\eeq
Other than $k_a$, the
only vector we have in this expression is $p_a$. The integral, evaluated with rotationally invariant
 cutoffs $\lambda$, $\mu$, will thus be proportional to
 $p_a$ and hence the expression in (\ref{B6}) will vanish.
 This confirms the assertion at the end of section 3 that
 the dissipation effect due to the soft photons
 will vanish.
 
There is another way to write down the solution
which will be useful for simplifying $V$. For this, without loss of generality,
we can take $Z^\mu (0)$, the origin of the world line of the droplet, to be zero. 
In the gauge $\del_\mu a^\mu = 0$,
the solution is given by
\beqar
a_\nu &=& - \int d^4y\,{d^4k \over (2\pi)^4} {1\over k^2 + i \epsilon} e^{-ik (x-y)} ~J_\nu (y)
\nonumber\\
&=& i  \int d^4y\,{d^3k \over (2\pi)^3} {1\over 2 \omega_k} \Bigl[
\theta(x^0 -y^0) e^{-i\omega (x^0-y^0) + i \vk\cdot(\vx-\vy)}\nonumber\\
&&\hskip 1.2in
- \theta(y^0 -x^0) e^{i\omega (x^0-y^0) - i \vk\cdot(\vx-\vy)}\Bigr] \, J_\nu (y)
\label{B7}
\eeqar
Carrying out the $y^0$-integration, we find, for $x^0 > y^0$,
\beq
a_\nu = q\, p_\nu \int {d^3k \over (2\pi)^3} {1\over 2\omega_k} 
~{1\over pk - i \epsilon} \, e^{i\vk\cdot {\tilde\vx}}
\label{B8}
\eeq
where $pk = p_0 \omega_k - \vp\cdot\vk$ and ${\tilde \vx} = \vx - (\vp x^0)/p_0$.
The contribution for $y^0 > x^0$ is similar except that we have 
$pk +i\epsilon$ in the denominator and
the exponential becomes $e^{-i\vk\cdot{\tilde\vx}}$. Combining the two and transforming to
the gauge $a_0 = 0$ and $\del_i a_i = 0$, we get
\beq
a_i = q \int {d^3k \over (2\pi)^3} 
{1\over 2 \omega_k} 
\left( \Omega_i^+ \,e^{-i\vk\cdot{\tilde\vx}} 
+ \Omega_i^- e^{i\vk\cdot{\tilde\vx}}  \right)
\label{B9}
\eeq
where
\beq
\Omega_i^{\pm}(k, p) = q \left( p_i - k_i {\vk\cdot\vp \over \vk^2} \right) {1\over pk \pm i \epsilon}
\label{B10}
\eeq
The corresponding electric field is given by
 \beq
 e_i = q \int {d^3k \over (2\pi)^3} 
{i \omega_k \over 2 \omega_k} 
\left(  \Omega_i^+ \,e^{-i\vk\cdot{\tilde\vx}}
- \Omega_i^-\, e^{i\vk\cdot{\tilde\vx}} \right)
\label{B11}
\eeq

The solutions (\ref{B9}) and (\ref{B11}) agree with
the results (\ref{B2}). To see this one needs to carry out the integrations
over $k_0$ and $\tau$  in (\ref{B2}).

\section*{Appendix C: Properties of expectation value for dressing factor}
\def\theequation{C\arabic{equation}}
\setcounter{equation}{0}

The function $W$ was defined in (\ref{48}) by
\beq
\bra{0,r} V(Q) \ket{0,r} = 
\bra{0,r} e^{\Phi}  \ket{0,r}= \bra{0,r} e^{i {\hat\chi}^A Q_A}  \ket{0,r}
\equiv e^{- {\half } W}
\label{C1}
\eeq
Here $\Phi = i {\hat \chi}^A Q_A$.
We will now show some of the properties of $W$ used in section 5.\\
{\underline{Property 1}}

W is real and $e^{-{\half}W } \geq 0$. This follows from the fact that we need even powers of
${\hat \chi}$ to obtain a nonzero Wick contraction.\\
{\underline{Property 2}}

$W$ is positive semi-definite.
The Cauchy-Schwarz inequality for states
$\ket{A} = e^\Phi \ket{0}$ and $\ket{B} = \ket{0}$ and
the fact that $(e^\Phi )^\dagger = e^{-\Phi}$, give the result
$\vert \la e^\Phi\ra\vert^2 \leq \bra{0} e^{-\Phi} e^\Phi \ket{0}$, 
which imples
$\vert \la e^\Phi \ra \vert \leq 1$; this shows that $W\geq 0$.
This is important in reconciling (\ref{53}) and (\ref{56}).\\
{\underline{Property 3}}

$W$ has the form
\beq
W = 2 \,Q^2  z \sum_0^\infty w_n \,\left( {z\,C_{\rm ad}  \over 2}\right)^n
\label{C2}
\eeq
Thus apart from the overall $Q^2$, the rest of the terms have the combination
$ {z\,C_{\rm ad} / 2}$. This result can be seen as follows.
First of all we carry out a scaling
$\Omega^\pm \rightarrow a\, \Omega^\pm$,
$Q_A \rightarrow Q_A/ a$ for some real parameter $a \neq 0$. Under this change,
$\Phi \rightarrow \Phi$, so $W$, defined in terms of $\Phi$ as in (\ref{C1}),
is unchanged.
The scaling is equivalent to
$z \rightarrow a^2 z$, $Q^2 \rightarrow Q^2/ a^2$.
The commutations rules for the $Q$'s become
\beq
\left[ {Q_A\over a} , {Q_B \over a} \right] = i {f_{ABC} \over a}
{Q_C \over a}
\label{C3}
\eeq
Thus, under the scaling, we must have
$f_{ABC} \rightarrow {f_{ABC} \over a}$ and
$C_{\rm ad} \rightarrow C_{\rm ad}/a^2$. Therefore, the invariant combinations are
$Q^2 z$ and $z C_{\rm ad} /2$, and $W$ must be a function of these.

Secondly, consider calculating $\la e^\Phi \ra$ by expansion of the exponential
for representations which are of large dimensions compared to the adjoint representation,
\beq
\la e^\Phi \ra = \sum_n {1\over (2n)!} \la \Phi \Phi \cdots \Phi\ra
\label{C4}
\eeq
After contracting the $c^A$, $c^{A\dagger}$ in $\Phi$, i.e., after using
(\ref{41}) and (\ref{49}), we get products like
$\la Q_A Q_B \cdots\ra$. The highest power in this corresponds to
$(Q^2)^n$ which is obtained by neglecting the commutators.
This contribution exponentiates to $\exp \left[ {1\over 2} \la \Phi \Phi\ra_c\right]$,
where the subscript $c$ indicates the connected component.
For large representations, the commutators are subdominant.
(This can also be seen by writing the product of the $Q$'s using star products.)
This fact shows that higher powers of $Q^2$ are excluded from $W$.
This can be seen by examining the behavior of a sample of two terms
in the expansion for $W$; the latter is of the form
\beq
{W\over 2} = {1\over 2} \la \Phi \Phi \ra_c + {1\over 4!} \la \Phi \Phi \Phi \Phi\ra_c + \cdots
\label{C5}
\eeq
where, again, the subscript $c$ denotes the connected component.
In $\la e^\Phi\ra$, the term with 6 $\Phi$'s is thus
\beq
\la e^\Phi\ra \sim {1\over 3!} ({\half} \la \Phi \Phi\ra_c)^3
+ ({\half} \la \Phi \Phi\ra_c) {1\over 4!} \la \Phi \Phi \Phi \Phi\ra_c + \cdots
\label{C6}
\eeq
The right hand side corresponds to the 6-point function for $\Phi$'s, which should be
of order
$(Q^2)^3$. This is already accounted for by
the $({\half} \la \Phi \Phi\ra_c)^3$ term. Thus if $\la \Phi \Phi \Phi \Phi\ra_c$ is
of any order higher than $Q^2$, we would have a contradiction. 
This shows that we can have only one overall power of $Q^2$ in $W$.
Thus, of the two invariants $Q^2 z$ and $z C_{\rm ad} /2$, we can have
only one power of $Q^2 z$ in $W$. The conclusion is that we do obtain the
result (\ref{C2}).

It is instructive to see how this works out for the first few terms of the expansion
of $\la e^\Phi\ra$. We can write this expectation value out as
\beq
\la e^\Phi\ra = 1 + {1\over 2} \la \Phi \Phi \ra + {1\over 4!}
\la \Phi \Phi \Phi \Phi\ra + \cdots
\label{C7}
\eeq
For the term with two $\Phi$'s, we get
\beq
\la \Phi \Phi \ra = - \la {\hat \chi}^A {\hat\chi}^B\ra \, \la Q_A Q_B \ra
= - z \la Q^2\ra
\label{C8}
\eeq
For the next term, we have via the Wick contractions
\beqar
\la \Phi \Phi \Phi \Phi\ra &=&
\wick{\c \Phi \c\Phi} \,\wick{\c \Phi \c\Phi}
+ \wick{\c1 \Phi \c2\Phi \c1\Phi \c2\Phi}
+\wick{\c1\Phi \c2\Phi \c2\Phi \c1\Phi}\nonumber\\
&=& 3\, \wick{\c \Phi \c\Phi} \,\wick{\c \Phi \c\Phi} + \left( 
\wick{\c1 \Phi \c2\Phi \c1\Phi \c2\Phi} - \wick{\c \Phi \c\Phi} \,\wick{\c \Phi \c\Phi}\right)
\label{C9}
\eeqar
The connected part of the 4-point function is thus given by
\beqar
\wick{\c1 \Phi \c2\Phi \c1\Phi \c2\Phi} - \wick{\c \Phi \c\Phi} \,\wick{\c \Phi \c\Phi}
&=& z^2 \la \left[ Q_A Q_B Q_A Q_B -
Q_A Q_A Q_B Q_B\right]\ra\nonumber\\
&=& - {C_{\rm ad} \over 2} Q^2 z^2
= - Q^2 z \left( {z C_{\rm ad} \over 2} \right)
\label{C10}
\eeqar
Notice that there is only one power of $Q^2 z$ here.
Proceeding in this way to the next order, we find
\beq
\la e^\Phi\ra =
e^{- W/2} =
\exp\left[ - Q^2 z \left( {1\over 2} + {1\over 4!} (z C_{\rm ad}/2) +
{5\over 6!} (z C_{\rm ad}/2)^2 + \cdots\right) \right]
\label{C11}
\eeq



\end{document}